\documentclass[doublecol]{epl2} 

\title{Estimate of the weight of empty space based on astronomical observations. }
\shorttitle{Estimate of the weight of empty space} 

\author{Thomas Durt\inst{1,2}}

\institute{                    
  \inst{1} Institut Fresnel, Domaine Universitaire de Saint-J\'er\^ome, Avenue Escadrille Normandie-Ni\'emen 
13397 Marseille Cedex 20, France.\\
  \inst{2} Ecole Centrale de Marseille, 38 rue F. Joliot-Curie, Technop\^ole de Chateau-Gombert, 13451 Marseille, France.
}

\abstract{
As a consequence of the equivalence principle and of the existence of a negative vacuum energy, we show how a weak but universal linear response of the vacuum to a local gravitational potential suffices to explain the main features of so-called Pioneer anomalies as well as the apparent departure from the  third Kepler law which has been observed at the level of numerous galaxies.}

\begin{document}

\maketitle

\section{Introduction}It is nowadays widely accepted that it is impossible to explain the deviations from Kepler's third law exhibited by stars at the level of numerous galaxies without formulating extra-hypotheses. The two main hypotheses that are presently \cite{SV,Blanchet1} competing are (1) the MOND theory and (2) the dark matter hypothesis. We propose a third approach (VIND for Vacuum Influenced Newtonian Dynamics) in which Newtonian dynamics is preserved while, in the absence of dark matter, it is the linear response of the vacuum energy which is supposedly responsible for apparent deviations to Kepler's third law. The paper is structured as follows. In the first section, we introduce the MOND, dark matter and (3) VIND models. In the second section we investigate the modifications to standard Newtonian dynamics deriving from our VIND model at small distances compared to the typical scale of the model ($L$). This allows us to estimate the value of $L$, on the basis of so-called Pioneer anomalies. We then investigate the modifications at large distances and obtain a good agreement between our predictions and MOND phenomenology. The last section is devoted to open questions, discussion and conclusion.
\section{MOND, dark matter and VIND models\label{0}}

(1) In the MOND theory  \cite{Milgrom} (``MOdified Newton Dynamics'') one assumes that Newton's law must be modified in a ad hoc fashion whenever we consider very small accelerations (relatively yo the inertial frame attached to ``fixed'' stars). It is motivated by the observed violations of Kepler's third law, which establishes that, as is the case in the solar system, celestial objects follow trajectories for which ``the square of the period $T$ varies like the cube of the mean distance $R$ to the sun''. In the case of circular orbits, this reflects Hooke's law: ${2\pi \over T}^2R=a^{Newton}$ is proportional to ${1\over R^2}$, up to constants. In other words, the Newtonian force $m\cdot a$ undergone by a test-particle of mass $m$ at distance $R$ of a ponctual source varies like $1/r^2$.  On the contrary, MOND theory in its simplest form predicts that the (intensity of the) acceleration undergone by a particle test in a very weak gravitational field (for instance far away from other gravitational sources) obeys
\begin{equation} |a^{MOND}|=\sqrt{a^{Newton}\cdot a_0},\label{mond}\end{equation} where $a_0$ is a new universal constant introduced by Milgrom. It is easy to check that, far away from a ponctual source of mass $M$, MOND theory predicts that \begin{equation}a=a^{MOND}=-\sqrt{G_N M\cdot a_0/r^2}=-(\sqrt{G_N M\cdot a_0})/r.\label{amond}\end{equation}

Therefore, the acceleration undergone by a test-particle at distance $R$ of a ponctual source varies like $1/R$ so that $({2\pi \over T})^2R$ is attractive and proportional to ${1\over R}$. In other words, the velocity $R/T$ is constant (does not depend on $R$) for any circular orbit, which is a fundamental property of MOND theories, that has been sometimes infirmed in the practice but has nevertheles been confirmed by numerous observations \cite{Tully,McGaugh} (this is sometimes called the flat velocity distribution).

(2) in the dark matter approach it is assumed that the departures from Kepler's third law are due to the presence of a huge concentration of dark matter, invisible to our telescopes and accelerators but responsible for the appearance of large accelerations which cannot be explained otherwise, if we stick to the Newtonian paradigm.

Both approaches present difficulties. The MOND approach can be considered as a clever and economic phenomenological description of what happens in a large number of galaxies, but it suffers from a clear theoretical background and motivation. For instance it is very difficult to conciliate MOND theories and general relativity, the latter predicting, roughly, that at slow velocities, Newtonian gravitation prevails.  It also suffers from clear experimental confirmations at the scale of clusters of galaxies and at the scale of the solar system.

The introduction of dark matter can also be considered to be an ad hoc hypothesis being given that, so far, all attempts to find a trace left by this hypothetical dark matter outside from the specific context of galactic rotations curves failed. At CERN for instance, none exotic particle described by dark matter theorists has been observed so far.

(3) The goal of the present paper is to show that the MOND predictions (or at least an ``average'' version of them) can be infered from an assumption that has the advantage to respect Newton dynamics. The price to pay is that we must assume that, everywhere in the cosmos, empty space exerts a weak back-reaction onto the gravitational potential. Otherwise, our results are derived in the framework of Newton's theory of gravitation. In particular we do not need to modify Newton's dynamical law ($F=ma$) at small accelerations, and the gravitational potential still evolves according to Poisson equation
\begin{equation} \Delta \phi=4\pi G_N \rho,\end{equation}
where $\phi$ is the gravitational potential, $G_N$ the Newton gravitational constant and $\rho$ the density of mass-energy. It has been recognised since several years that in the Newtonian limit the density of mass-energy ought to contain a contribution due to the so-called cosmological constant, which leads to the modified Poisson equation \cite{poincar}
\begin{equation} \Delta \phi=4\pi G_N (\rho^{matter}-2\rho_\lambda),\label{modP3}\end{equation}
where $\rho_\lambda=\lambda/8\pi G_N$ and $\lambda$ is Einstein's cosmological constant. It is not completely clear whether this contribution is due to quantum vacuum fluctuations or not but here we shall take for granted that empty space possesses a negative mass density, constant in space.

The novelty of our approach is that we also assume that, as a consequence of the mass-energy equivalence principle, the matter and the vacuum do not only contribute to the density of mass-energy as isolated source terms, but that they also exert a back reaction proportional to the gravitational field:
\begin{eqnarray} \Delta \phi=4\pi G_N(\rho^{matter}-2\rho_\lambda)(1+\phi/c^2)\nonumber \\=4\pi G_N(\rho^{matter}-2\rho_\lambda)+(4\pi G_N c^{-2}\rho^{matter}-k^2)\phi,\label{modP},\end{eqnarray}
where $k^2=8\pi G_Nc^{-2}\rho_\lambda=\lambda c^{-2}$.

We are not interested here in computing effects due to the linear in $\phi$ back reaction of matter ($4\pi G_N c^{-2}\rho^{matter}\phi$). They can be evaluated independently and in general they only result in a small correction to usual Newtonian dynamics. For instance, at the level of the solar systems where the gravitational potential is everywhere small relatively to $c^2$, they does not modify much the Newtonian, zero-order dynamics. Nevertheless, the back reaction due to the vacuum energy, despite of the fact that it is weak, is present everywhere in space and its cumulative effects may not be neglected. Our goal is to show how these effects provide a plausible explanation of the observed deviations to Kepler's third law, and of so-called Pioneer anomalies as well...

In summary, Vacuum Influenced Newton Dynamics derives from the assumption that Newton dynamics is not modified but ``only'' indirectly influenced by the appearance of a new source term $k^2\phi$ that we ultimately attribute to vacuum\footnote{See ref.\cite{Blanchet2} for a similar angle of approach.}. 
\section{VIND in the near and far field regimes \label{2}}Formally, the equation (\ref{modP}), in vacuum, can be solved once we know the associated Green's function, which is a solution of Green's equation
\begin{eqnarray} \Delta G({\bf x},{\bf x'})+k^2G({\bf x},{\bf x'})=\delta ({\bf x}-{\bf x'}),\label{green}\end{eqnarray}
where $\delta ({\bf x}-{\bf x'})$ represents the three-dimensional delta function. As is well-known, even if we impose that the Green function is isotropic (that is to say, it depends only on the distance between  ${\bf x}$ and ${\bf x'}$), it is not unique in the sense that all Green functions differ by a solution of the homogeneous Helmholtz equation. We are interested in a real and isotropic Green function, which suggests the choice \begin{equation}G({\bf x}-{\bf x'})={-cos(k(|{\bf x}-{\bf x'}|)\over 4\pi |{\bf x}-{\bf x'}|}\label{greenreal}.\end{equation} We are free to add to the Green function  defined by equation (\ref{greenreal}) a real multiple of the harmonic function ${sin(k(|{\bf x}-{\bf x'}|)\over 4\pi |{\bf x}-{\bf x'}|}$. However, in realistic situations, the weight of such a solution remains very small, and the essential conclusions of all our analysis remain unchanged.

In particular, convoluting the omnipresent vacuum energy $-2\rho_\lambda$ everywhere in space with the Green function yields a finite constant additive correction to the potential. The contribution in a location ${\bf x}$ essentially comes from a region situated at a finite number of wavelenghts (denoted $L$, $L=2\pi/k$) around ${\bf x}$. It is not necessary to make particular assumptions about the structure of the universe (e.g. is it finite, open or else) in order to renormalise properly the gravitational potential in the present case, excepted that $L$ supposedly differs from 0. We shall thus consistently forget in what follows the presence of the source term $-2\rho_\lambda$ in equation (\ref{modP}) and focus on the changes brought by the linear vacuum response $k^2\phi$.

\subsection{Near field regime}At small distances ($|{\bf x}-{\bf x'}|<<L$), the Green function reduces at the zeroth order to the usual Green function of Newton's theory ${-1\over 4\pi|{\bf x}-{\bf x'}|}$, which means that our model is nearly equivalent to Newton's model for gravitation at small distances from the sources of matter. As it is granted that, in the solar system, Newton's model works very well\footnote{Up to the so-called Pioneer anomalies \cite{PRA} that we discuss soon, and up to relativistic corrections which, anyhow, remain out of the scope of the present work, merely aiming at describing the motion of slowly moving objects relatively to ``fixed stars''.}, we must impose that $L$ is quite larger than the size of the solar system. It is however interesting to take account of slight but systematic deviations of Newton's theory which have been observed with slow objects (the Pioneer 10 and 11 probes), and constitute what is called the ``Pioneer anomaly''.

In short, it has been repeatedly observed at the level of these probes (twice) that a constant centripetal acceleration seemingly prevails in our solar system. This acceleration is constant in time and space and directed towards the sun. Its magnitude is  \cite{PRA} of the order of $8\cdot 10^{-10}$ m/s$^2$. 

Our model precisely predicts such an effect, as can be seen by developing the Green function (\ref{greenreal}) to the lowest order in $k$ which reads

\begin{equation}G({\bf x}-{\bf x'})\approx{-1\over 4\pi |{\bf x}-{\bf x'}|}+{k^2\over 8\pi}|{\bf x}-{\bf x'}|\label{greenreal}.\end{equation}
Sufficiently far from the sun (where ${\bf x}-{\bf x'}$ is in good approximation constant in magnitude and direction for all source terms inside the sun (such that $|{\bf x'}|<R_{sun}$)), so that we expect that a constant radial and centripetal force of intensity $G_N\cdot M_{sun}\cdot k^2/2$ must exist, which adds to the usual Newtonian solar attraction. Identifying this force with the Pioneer anomaly, we derive an estimate of $k$: $k^2=2a_{Pioneer}/(G_N\cdot M_{sun})$. Injecting in this relation the typical values $M_{sun}\approx 2 \cdot 10^{30}$ kg, $a_{Pioneer}\approx 8 \cdot 10^{-10}$ ms$^{-2}$ and $G_N\approx 7\cdot 10^{-11}$ m$^3$ s$^{-2}$ kg, we find

$k=2\pi/L\approx 3\cdot 10^{-15}$m$^{-1}$. This provides an estimate for $L$: \begin{equation}L\approx 2\cdot 10^{+15}\rm{m}\end{equation}  One light year being equal to $9,461 \cdot 10^{15}$ m, we find that $L$ is more or less one fifth of a light year. This justifies, a posteriori, why the Pioneer anomaly is so weak while the usual Newtonian theory is so efficient at the scale of the solar system.

 These results are in agreement with the constraints derived so far, which are:

(1) $L$ differs from 0.

(2) $L$ is quite larger than the size of the solar system.

Actually, although a more detailed treatment is out of the scope of the present paper, it is not too difficult to convince oneself that, in the approximation where we treat the sun as a sphere where $\rho^{matter}$ is taken to be constant and equal to $<\rho^{matter}>_{sun}$, it is relatively easy to estimate the perturbation resulting from (the modification of the usual Newton-Poisson equation by) new source terms including the vacuum density, as well as the back reaction of matter and vacuum resulting from the equivalence principle (as described at the level of equation (\ref{modP})). This refined model leads thus to the same prediction as before, which is, it predicts as main perturbation at the level of the solar system the existence of a constant acceleration directed towards the sun, of magnitude $G_N\cdot M_{sun}\cdot k^2/2$.

\subsection{Coherent  interferences in the far field regime}

If we consider a distribution of matter of size smaller than or comparable to the size of the solar system (like a typical star for instance), the potentials generated by each element of the matter distribution will coherently interfere in the ``far field regime'' ($|{\bf x}-{\bf x'}|>>L$). A striking novelty of our approach, compared to the ``standard'' Newtonian approach is that, in this case, the  gravitational field generated by a mass $M$ located at the position ${\bf x}'$,  at the position ${\bf x}$, is no longer the Newton-Coulomb potential ${-G_NM\over |{\bf x}-{\bf x'}|}$, but it is the oscillating function ${-G_NMcos(k(|{\bf x}-{\bf x'}|)\over  |{\bf x}-{\bf x'}|}$. This leads to several counterintuitive properties, the first of them being that gravitational forces are not necessarily attractive.  Indeed, the acceleration $a^{VIND}$ generated by a mass located at position ${\bf x'}$ onto a test-particle located at position ${\bf x}$ obeys \begin{eqnarray}a^{VIND}={-\partial \over \partial r}({-G_NMcos(k(|{\bf x}-{\bf x'}|)\over 4\pi |{\bf x}-{\bf x'}|})\nonumber\\= {-kG_NMsin(k(|{\bf x}-{\bf x'}|)\over  |{\bf x}-{\bf x'}|}-{G_NMcos(k(|{\bf x}-{\bf x'}|)\over  |{\bf x}-{\bf x'}|^2}\label{vind}\end{eqnarray}
Therefore, in the far field regime (at distances larger than one light year), the dominating term of the potential created by any celestial object of size comparable to or smaller than the size of the solar system is, generically, the first one in the last part of equation (\ref{vind}):
\begin{equation}a^{VIND}\approx {-kG_NMsin(kr)\over  r}\label{vind2}\end{equation}

The expression (\ref{vind2}) obviously differs from the expression (\ref{amond}). Now, as we have shown in the previous section, $L$ being smaller than one light year, it is significatively smaller than typical galactic sizes (even smaller than typical sizes of galactic cores), which implies that the potential generated by the stars of the galactic core will not always constructively interfere. As we shall now show, it is possible to find a good agreement between MOND and VIND models in this case (incoherent interferences in the far field regime), provided we interpret the MOND theory in a statistical sense.
\subsection{Incoherent interferences in the far field regime}

Another counterintuitive feature of our model is that gravitational potentials generated by different sources of matter do not always add to each other in a simple way but have the ability to interfere with each other. This property is obvious if we consider electro-magnetic potentials but is totally absent from the Newtonian paradigm. To illustrate the consequences of this property, let us assume that the core of a galaxy is densely populated so that the law of large numbers is valid.
The acceleration generated by the core of the galaxy onto a test particle (or another star) located in the suburbian zone of the galaxy will then obey (where the symbol ...$\approx$... must be understood here as...is equal ``in average'', or is ``typically'' equal to...):
\begin{equation}|a^{VIND}|\approx {-\tilde C\sqrt{N_\odot } G_N k<M_\odot >\over  r}\label{vind3},\end{equation}where $\tilde C$ is a constant of the order of 1 while $<M_\odot >$ is the average mass of a star of the galactic core and $N_\odot$ its number of stars. It must be so because the phases which characterize the potentials generated by the different local clusters formed by the $N_\odot $ stars of the core will add incoherently, so that the average ``phase factor'' $<\sum_{i=1}^{N_\odot}sin(kr_i)>$ obtained after summing over their incoherent contributions will be Gaussianly distributed around zero, with a variance $\sigma$ equal to $\tilde C \cdot\sqrt {N_\odot } $, with $\tilde C$ a number of the order of unity\footnote{From now on we shall make the natural choice $\tilde C=\sigma(sin(kr))\approx \sqrt{1/2}=\sqrt{\int_0^{2\pi}sin^2xdx/2\pi}$)}, as predicted by the law of large numbers.

Consequently, $a^{VIND}$, the (averaged over space) value of the acceleration  towards the core of the galaxy, predicted in the VIND approach, will be statistically distributed around the value 0, with a spread of the order of ${\sigma \cdot k G_N <M_\odot >\over  r}$=$\tilde C{\sqrt {N_\odot}G_N k<M_\odot >\over  r} $, as announced at the level of equation (\ref{vind3}).

We find in this case that the ratio between $<a^{VIND}>$ and $a^{MOND}$ obeys

\begin{equation}{|a^{VIND}|\over a^{MOND}}\approx \tilde Ck\sqrt{{G_N<M_\odot >\over a_0}}\label{ratoint}=k\sqrt{{G_N<M_\odot >\over 2a_0}}.\end{equation} 

Injecting in equation (\ref{ratoint}) the typical values $M_{sun}\approx 2 \cdot 10^{30}$ kg, $a_0\approx 1,2 \cdot 10^{-10}$ ms$^{-2}$ and $k\approx 3\cdot 10^{-15}$ m$^{-1}$, we find that

\begin{equation}{|a^{VIND}|\over a^{MOND}}=\sqrt{{a_{Pioneer}\over a_0}}\sqrt{{<M_\odot >\over M_{sun}}}.\approx 2,5\sqrt{{<M_\odot >\over M_{sun}}}.\label{calib}\end{equation}

This demonstrates how, in a statistical sense, MOND phenomenology can be derived from the assumptions that we originally formulated about vacuum energy, which are the basis of our VIND model.

That this ratio is not exactly equal to one is not amazing for two reasons:

(1) MOND's acceleration merely provides an order of magnitude. Indeed, in the framework of MOND's phenomenology it is well-known that one free parameter must be adjusted, for each galaxy, in order to take account of the mass of the galaxy.

(2) the ratio $\sqrt{{a_{Pioneer}\over a_0}}$ which appears in equation (\ref{calib}) is likely to increase due to various sources of decoherence and to decrease due to the presence of dark positive matter-energy. For instance, due to the spread of $k$ one should expect that the effective MOND acceleration diminishes with the distance to the center of the galaxy, due to decoherence at large distances.

A fine study of the calibration factors which are introduced galaxy by galaxy provide a possible test aimed at discriminating VIND and MOND theories. In particular the scaling in terms of $\sqrt{{<M_\odot >\over M_{sun}}}$ is not the same in both approaches\footnote{MOND's acceleration scales like $(<M_\odot >)^{1/2}$, while VIND's acceleration scales like $(<M_\odot >)^{1}$.}. Other tests are discussed in the last section.

\section{Open questions, discussion and conclusion \label{3}}\subsection{Other possible experimental tests}We mentioned three possible tests which confirm or could confirm that our ideas are well-founded (Pioneer anomalies, flat galactic velocity distributions and fine study of the calibration parameters and velocity curves), for which no new brutto experimental data must be collected \cite{PRA,Tully,McGaugh}. Other tests could be realized in the futureas we discuss now.

Although both theories predict that the direction of the acceleration undergone by a test mass (for instance a star in the suburbs of a galaxy) is directed towards the centre of the galaxy, the VIND and MOND approaches differ by an essential feature: the intensity of this acceleration slowly varies over space in the MOND model while, in the VIND model, it undergoes a strong relative change (it can even change its sign) at a scale $L$ (supposedly of the order of a fifth of light year). 
This implies another main difference between both theories: MOND theory is {\it per se} a deterministic theory while, due to the random distribution of constructive and destructive interference effects over space, VIND predictions are {\it per se} statistical. It is only in average that both theories lead to comparable predictions. This suggests several approaches aimed at discriminating between both explanatory schemes. The first of them consists of checking the sign of the centripetal acceleration undergone by stars located at the edge of dense galaxies, but this is not an easy task for experimentalists because we are dealing here with tiny velocities and accelerations. In the case of nearly circular orbits around the galactic core, the radial velocities are particularly small and vibrations along the radial axis predicted by the VIND model should be particularly slow.

Now, even if the results confirm that the sign of the centripetal acceleration is nearly always positive, this does not necessarily mean that the VIND approach must be rejected. The reason is that the distribution of the trajectories of stars, in the VIND approach, is not necessarily homogeneously spread over space. Indeed, it is reasonable to suppose that the zones where  gravitation is repulsive (roughly regions where $<sin(kr)>$ is negative) are unstable because there the test particle undergoes a repulsive gravitational interaction which can not be counterbalanced by the centrifugal force. It is therefore likely that the regions where $sin(kr)$ is negative are sparsely populated (which is confirmed by the fact that a flat velocity distribution has been observed: when the gravitational force is repulsive the distance to the galactic core and the tangential velocity are a priori uncorrelated). The latter prediction, in turn, is likely to be submitted to experimental observations because it would imply that inside a given galaxy populated regions of extent $L$ would coexist with empty regions of comparable size. Actually, in order to derive the distribution of trajectories of stars of a given galaxy, in the VIND approach, one must ultimately describe the full history of these trajectories which is a very difficult task. Nevertheless, it is very unlikely that this history conspires to reproduce the deterministic predictions of MOND theory and therefore a measure of statistical fluctuations of the velocities of stars situated at the same distance of the core of a galaxy should suffice in principle to discriminate between MOND and VIND predictions.

Remark that the VIND theory also predicts the appearance of repulsive barrers around a star, at distances separated by integer multiples of $L$, which could maybe be revealed by a direct study of accretion disks of stars present in media where a lot of interstellar gas is present. This would be the firt experimental evidence of negative (repulsive) gravitational interaction.

Another situation of interest occurs when the law of large numbers is not valid, for instance when a small number of stars or a small number of galactic clusters are present. This is interesting because it is known that in this regime the MOND approach does not fit well with observations. Let us consider for instance that three stars are aligned, and separated by a distance larger than $L$, and let us treat the star situated in the middle as a test particle. In the MOND approach it will undergo an acceleration  $\sqrt{G_N (M_1/r_{13}^2-M_2/r_{23}^2)\cdot a_0}$. In the VIND approach however it will undergo an acceleration equal to $|{-G_NM_1sin(kr_{12})\over  r_{12}}+{G_NM_2sin(kr_{13})\over  r_{13}}|$. Obviously these predictions have few in common.

\subsection{Conclusion}

Before we end this last section, it is worth noting that our model indirectly provides an estimate of the cosmological constant $\lambda=c^2\cdot k^2\approx (6\pi)^2 \cdot10^{-16} \approx 4\cdot10^{-14}$s$^{-2}$, to compare with the  lower value of $10^{-35}$s$^{-2}$, which has been derived on the grounds of rather recent astronomical observations \cite{Perl} (aimed at measuring indirect manifestations of the existence of a non null cosmological constant predicted in the framework of general relativity), and with the larger value of $10^{87}$s$^{-2}$ which derives from a cut-off at the Planck scale \cite{poincar}. Despite of the fact that it is not clear today, to anybody, what is the meaning of the cosmological constant\footnote{Averaging the acceleration expresssed in equation (\ref{vind}) over a spatial period, we see that gravitation vanishes. Rather ironically, although Einstein introduced the cosmological constant in order to prevent \cite{poincar} a scenario of the big bang type, the effect of a constant negative mass is seen, in our approach, to prevent a big crunch scenario.}  and how and why it should be related to quantum vacuum fluctuations and quantum vacuum energy, we were pleased to elaborate on the idea that a negative and constant ``etheric'' distribution of mass is omnipresent in the cosmos. As we have shown, the recognition of this fact has implications at the level of astronomy which cannot be ignored, and it is not necessary to resort to sophisticated theories (like general relativity, exotic particles and so on) in order to figure out what some of these implications could be.

The most amazing lesson that we can draw from our analysis is that (in the case where our theoretical building would appear to resist to facts) it allows us to infer ``the weight of empty space''. Indeed, it predicts for the vacuum density of mass $-2\rho_\lambda=-k^2c^{2}/4\pi G_N$ a value close to ${\bf -} 0,9 \cdot 10^{-3}$ kg/m$^3$, more or less MINUS one gram per cubic meter...Once we opened the Pandora box, many questions spontaneously arise, e.g., ``is $k$ constant throughout time ?'', ``to which extent does its existence influence the cosmologic evolution (expansion, galactogenesis, distribution of the cosmic background and so on) ?'', ``can its value be infered from the standard particle model ?''...As we see, black clouds in the blue sky of astrophysics possibly generate rain and rainbows.

\acknowledgments
As an outsider to the field who essentially built his culture of the topic through the reading of magazines, wiki sites (e.g. the site of John Baez dedicated to vacuum's energy (http://math.ucr.edu/home/baez/vacuum.html), the wiki sites respectively dedicated to the cosmological constant and the MOND theory) and so on, the author wishes to apologize in advance in the case where the bibliography would be incomplete.

\end{document}